\documentclass[10pt]{emulateapj}
\usepackage{epsfig}
\usepackage{amsmath}
\usepackage{multirow}
\usepackage{multirow}

\begin{document}

\title{Energy spectrum and chemical composition of ultra-high energy cosmic rays from semi-relativistic hypernovae}

\author{Ruo-Yu Liu\altaffilmark{1,2} and Xiang-Yu Wang\altaffilmark{1,2}}
\altaffiltext{1}{Department of Astronomy, Nanjing University,
Nanjing, 210093, China} \altaffiltext{2}{Key laboratory of Modern
Astronomy and Astrophysics (Nanjing University), Ministry of
Education, Nanjing 210093, China}

\begin{abstract}
It has been suggested that hypernova remnants, with  a substantial
amount of energy in semi-relativistic ejecta, can accelerate
intermediate mass or heavy  nuclei to ultra-high energies and
provide sufficient amount of energy in cosmic rays to account for
the observed flux. We here calculate the expected energy spectrum
and chemical composition of ultra-high energy cosmic rays  from such
semi-relativistic hypernovae. With a chemical composition  equal to
that of the hypernova ejecta and a flat or hard spectrum for cosmic
rays at the sources, the  spectrum and composition of the propagated
cosmic rays observed at the Earth can be compatible with the
measurements by the Pierre Auger Observatory.

\end{abstract}

\keywords{ cosmic rays --- gamma rays: bursts }

\section{Introduction}
Ultra high energy cosmic rays (UHECRs) are the most energetic
particles ever detected in the universe. There is a  consensus that
UHECRs with energies larger than $10^{19}$eV are of extra-galactic
origin. However, the composition of ultra high energy cosmic rays at
these energies remains disputed. Recent observations of their
maximum air shower elongations ${\rm <X_{max}>}$ and fluctuations
${\rm RMS(<X_{max}>)}$ by the Pierre Auger Observatory (PAO) suggest
that UHECRs are progressively dominated by heavier nuclei at
energies above a few EeV \citep{Abraham10b}, while { the results
from High Resolution Fly's Eye Experiment (HiRes) and Telescope
Array (TA) are consistent with a pure  proton
 composition \citep{Abbasi04, Tsunesada11}}.  These claims, however,
depend on the poorly-understood hadronic interaction models at such
high energies. Nevertheless,
the advantage of heavy composition for UHECRs is that
heavy or intermediate-mass particles are much easier to be
accelerated to UHE energies than protons as the maximum particle
energy is rigidity dependent (i.e. proportional to the nuclear
charge $Z$ of the particles). Heavy nuclei with energy above { 100}
EeV will suffer from strong photo-disintegration attenuation due to
interactions with cosmic microwave background (CMB) and cosmic
infrared background (CIB) photons, with an attenuation length
shorter than $\sim 100$ Mpc (see e.g. \citealt{Puget76, Allard06,
Hooper07, Allard09, Aloisio11, Taylor11}). Thus, sources producing
UHECRs above  100 EeV must  be within $\sim100$ Mpc. Within this
distance, there are few sources that are powerful enough to be able
to accelerate particles to energies $10^{20}$ eV. The candidates
include local active galactic nuclei (AGNs) \citep[e.g.][]{Biermann87,
Berezinsky06,Peer09} and gamma-ray bursts (GRBs) \citep[e.g.][]{Waxman95,Waxman04,Vietri95,Wick04,Dermer06,Murase06a,Murase06b}.

Extragalactic hypernova has been suggested to be a candidate source
of UHECRs as well\citep{Wang07,Wang08, Liu11}. Hypernovae are a
peculiar type of supernovae that are more bright and have larger
explosion energies than typical ones \citep{Paczynski98,
Soderberg06}. Several hypernovae have been observed in the nearby
universe to date, usually associated with sub-energetic GRBs. For
example, SN 1998bw, associated with GRB 980425, is the first-found
peculiar supernova at a distance of 38 Mpc, with an
isotropic-equivalent total kinetic energy of $\sim 5\times
10^{52}$erg \citep{Galama98, Kulkarni98, Iwamoto98}.  Radio
afterglow of this event showed that more than $10^{50}$  erg of
kinetic energy was released in the form of a mildly relativistic
ejecta.   SN 2006aj, associated with GRB 060218, was detected at a
distance of 140 Mpc
\citep{Campana06,Mirabal06,Soderberg06,Mazzali06b}. The strong
thermal  X-ray emission component in GRB 060218 is interpreted as
arising from  a semi-relativistic supernova shock breakout, in which
a mildly relativistic ejecta with energy about $10^{50}$ erg has
been inferred \citep{Campana06, Soderberg06}. Radio observations of
the recently discovered supernova SN2009bb, for which no associated
GRB  is found, also suggest that more than $10^{50}{\rm erg}$ is
coupled into the mildly-relativistic ejecta\citep{Soderberg10}. We
will use the term semi-relativistic hypernovae to denote such
supernovae harbouring mildly relativistic ejecta.  Such
semi-relativistic ejecta  drives semi-relativistic shocks, which
could  accelerate particles to ultra-high energies \citep{Wang07}.
The recently proposed engine-driven supernova as a source for UHECRs
are identical to the hypernovae in nature \citep{Ray11}, since both
are powered by the central engine and both have much more energy in
semi-relativistic ejecta. The estimated event rate of SN2009bb-like
engine-driven supernovae is also comparable to the hypernova
rate\citep{Soderberg10}. The progenitors of hypernovae are believed
to be Wolf--Rayet (WR) stars, which are stripped of their original
H--rich and sometimes even He--rich envelopes. The circum--stellar
winds are therefore rich in intermediate mass elements, such as C
and O.  Moreover,  the hypernovae ejecta contain heavy elements such
as Fe that are produced by nuclear fusion in the core of exploding
WR stars, as well as elements heavier than Fe that are synthesized
during the explosions.

In this paper, we study the propagated energy spectrum and chemical
composition of UHECRs that originate from hypernovae, and confront
them with the measured data by PAO. The rest of this paper is
organized as follows. In \S 2, we first briefly review the hypernova
scenario and describe the source  composition of UHECRs produced by
hypernovae. In \S 3, we calculate the energy spectrum and
composition of UHECRs arriving at the Earth after propagation and
then confront them with the Pierre Auger observations. Finally, we
give our discussions and conclusions in \S 4. The effect of
intergalactic magnetic fields is neglected in this paper. Throughout
the paper, we use eV as the unit of particle energy and use c.g.s
units for other quantities and denote by $Q_x$ the value of the
quantity $Q$ in units of $10^x$.

\section{The  chemical composition and energy spectrum of UHECRs at the source}
\subsection{Hypernova model for UHECRs}

\citet{Wang07,Wang08} have suggested that hypernova remnants can
accelerate heavy or intermediate mass nuclei to ultrahigh energies
above $10^{19}{\rm eV}$. Here for simplicity we only focus on the
semi-relativistic part of the ejecta, since only this part is
relevant to the acceleration of UHECRs. Particles are accelerated in
the shock region where the semi-relativistic ejecta is freely
expanding before being decelerated by the swept-up circum-stellar
medium.  The size of this free-expansion phase region for ejecta of
a particular velocity $\beta_{\rm sh}c$ and kinetic energy $E_K$ is
$R_{\rm HN}=E_k v_w/\Gamma^2c^2\dot{M} \simeq 2\times10^{17}
E_{k,51} \Gamma^{-2} \dot{M}_{-5}^{-1} v_{w,3} {\rm cm}$, where
$\Gamma=(1-\beta_{\rm sh}^2)^{-1/2}$ is the bulk Lorentz factor of
the ejecta, ${\dot{M}}=10^{-5}\dot{M}_{-5}{\rm M_\odot yr^{-1}}$ is
the wind mass loss rate (whose average value is $3\times10^{-5} {\rm
M_\odot yr^{-1}}$ for WR stars), and $v_w = 10^3 v_{w,3}\rm
kms^{-1}$ is the wind velocity \citep{Willis91, Chevalier99}.  We
assume a fraction of $\epsilon_B$ of shock internal energy goes into
the magnetic field and hence the magnetic field energy density is
$B^2/8\pi \approx 2\Gamma^2\beta^2\epsilon_B \rho_w c^2$, where
$\rho_w=\dot{M}/4\pi R_{\rm HE}^2v_w$ is the mass density of the
stellar wind at radius $R_{\rm HN}$. Thus, by equating the adiabatic
cooling time with the acceleration time, the maximum energy of
accelerated particles is
\begin{equation}\label{Emax}
\begin{split}
E_{\rm max}\simeq& Z e \Gamma BR_{\rm HE}\beta_{\rm sh} \\
=&3.7\times10^{20}  \left(\frac{Z}{26}\right)
\epsilon_{B,-1}^{1/2}\Gamma^2\beta_{\rm sh}^2\dot{M}_{-5}^{1/2}
v_{w,3}^{-1/2} {\rm eV}.
\end{split}
\end{equation}
where  $Z$ is the nuclear charge number of the particle and $e$ is
the charge of electrons. { Note that, in the semi-relativistic
hypernova scenario, the adiabatic cooling of UHE nuclei is much more
efficient than other cooling processes, such as synchrotron cooling,
photopion production  and photo-disintegration processes
\citep{Wang08}, so we do not consider them here.}

{Hypernova can also provide sufficient energy in
UHECRs{\citep{Wang07, Liu11}. According to \citet{Katz09}, the
energy production rate\footnote{ The value obtained by
\citet{Katz09} is based on pure-proton composition assumption. Since
iron nuclei attenuation length around $10^{20}$eV is similar to that
of proton, we adopt this value as the energy production rate in pure
heavy nuclei scenario.} in UHECRs  per logarithmic energy interval
inferred from the measured flux  by PAO  is $\sim
10^{43.5}$ergMpc$^{-3}$yr$^{-1}$. Hypernova ejecta with $\Gamma\beta
\sim 0.5$ can produce cosmic rays with energies $\sim10^{20}$eV and
the kinetic energy in such velocity ejecta is typically
$\sim10^{51}$erg \citep[e.g.][]{Wang07,Chakraborti11}. Assuming that
half of the kinetic energy goes into cosmic rays and that $10\%$ of
this energy is distributed in  every logarithmic energy interval,
the required local event rate of hypernovae is then
$\sim600$Gpc$^{-3}$yr$^{-1}$, which is consistent with the observed
rate (e.g. Soderberg et al. 2006; Liang et al. 2007). }

\subsection{Composition and spectrum at the sources}

It is usually assumed that shock acceleration leads to a power-law
energy spectrum for particles. The {differential} number of
particles accelerated by one hypernova can be described by
\begin{equation}
\frac{dN_i}{dE_i}=\frac{U_iE_i^{-p}}{\int_{E_{i,{\rm
min}}}^{E_{i,{\rm max}}}dE_i E_i^{1-p}} {\rm
exp}(-\frac{E_i}{E_{i,\rm max}}){\rm exp}(-\frac{E_{i,\rm min}}{E_i})
\end{equation}
where the subscription $i$ represents  the species of the nuclei
(e.g. $i$=H, He, and etc.), $E_{i,\rm max}$ and $E_{i,\rm min}$ are
respectively the maximum and minimum energy of cosmic rays injected
into the intergalactic space by the hypernova,  and $U_i$ is the
total energy of nuclei of species $i$ released by one hypernova into
the intergalactic space. According to Eq.~\ref{Emax}, hypernova
remnants can accelerate iron nuclei to energies about $\sim5\times
10^{20}$eV, so we fix $E_{\rm max, Fe}=10^{20.5}(Z/26)$eV in the
following calculation. $E_{i,\rm min}$ is the low energy cutoff in
the spectrum, since nuclei with energy below $E_{i,\rm min}\approx
Z_ieB_{\rm host}R_{\rm host}$ can not escape from the confinement by
the magnetic fields in the host galaxies, where $B_{\rm host}$ is
the typical magnetic field of the host galaxy and $R_{\rm host}$ is
the scale height of the host galaxy. Taking our own galaxy as an
approximation, we have $E_{\rm min, Fe}\approx
10^{19}(Z/26)B_{-6}(h_{z}/0.3\rm kpc)$eV, where $h_z$ is the scale
height of the Galaxy. For a single velocity ejecta \footnote{It has
been suggested that hypernova produces a kinetic energy distribution
spreading over a range of velocity as $E_k\propto
(\Gamma\beta)^{-\alpha}$ (e.g. \citealt{Soderberg06}). The final
injected CR spectrum from the hypernova should be contributed by
different velocity ejecta  and the superimposed spectrum becomes
softer\citep{Wang07}. {We here consider the highest energy cosmic
rays, and for simplicity, just focus on the fastest
(semi-relativistic) part of the ejecta and  assume $p\simeq2$ in the
following calculation.}}, the power--law index is usually assumed to
be $p\simeq2$ for both non-relativistic shocks and semi-relativistic
shocks \citep{Kirk87}. In the following calculation, we will assume
the spectrum to be power--law with exponential cutoffs above
$E_{\rm max}$ and below $E_{\rm min}$.

Given that both $E_{i,\rm max}$ and $E_{i,\rm min}$ are rigidity
dependent  (i.e. $E_{i,\rm max(min)}\propto Z_iE_{\rm H,max(min)}$),
the total mass of nuclei of species $i$ released into the
intergalactic space by one hypernova is
\begin{equation}
M_{i,\rm CR}=A_i\int_{E_{i,\rm min}}^{E_{i,\rm
max}}\frac{dN_i}{dE_i}dE_i\propto \frac{A_i}{Z_i}U_i
\end{equation}
Approximating $Z_i=A_i/2$ (except for H), and assuming that the same
proportions of  particles of each species  get accelerated (i.e. the
value of $M_{i,\rm CR}/M_i$ is independent of $i$),  we have
\begin{equation}
U_{\rm H}:U_{i}:U_{j}= 2M_{\rm H}:M_{i}:M_{j}.~~~~~(i,j\neq H, i\neq
j)
\end{equation}
With the above equation, one can transform the abundances of nuclei
of each species to the abundances  of these nuclei at certain given
energy, as described by Eq.(2).

In the hypernova scenario, cosmic ray particles originate from the
circum-stellar wind material or hypernova ejecta material. The WR
stellar wind contains abundant intermediate mass elements such as C
and O, which dominate the chemical enrichment of interstellar medium
(ISM) (e.g. \citealt{Abbott82,Bieging90}). { The mass ratio of
different elements for  typical WR stellar wind is $M_{\rm He}:
M_{\rm C}: M_{\rm O}: M_{\rm X}=0.32: 0.39: 0.25: 0.04$ \citep{Bieging90}, derived
based on observation of emission line of these elements in WR stars
and the stellar evolution model (see e.g. \citealt{Willis82,Prantzos86,Hucht86}, where X denotes
the elements other than He, C and O.} In the following calculation,
we will use this composition as a representative case for the cosmic
rays from the stellar wind.

On the other hand, the hypernova ejecta contains both intermediate
mass elements and heavy elements. It consists of products of nuclear
reaction in the interior of the WR star, such as C, O, Mg, Fe, as
well as some heavier elements produced by the explosive
nucleosynthesis. Chemical composition of the ejecta can be derived
by modeling the spectra and light curves of hypernovae (e.g.
\citealt{Nakamura01, Fryer06}) or through numerical simulation of SN
explosion based on some specific stellar model (e.g.
\citealt{Georgy09}). Best fit for the early spectra and light curve
of SN 1998bw is obtained in the model named "CO138E50" by
\citet{Nakamura01}, where the input explosion kinetic energy and
ejecta mass are, respectively, $E_K=5\times 10^{52}$ergs and $M_{\rm
ej}=10M_{\odot}$. This model yields a chemical
composition\footnote{We neglect the elements with mass less than
$10^{-3}M_{\odot}$ and treat elements heavier than Fe (e.g. Ni, Co)
as Fe because such elements would decay into Fe finally.} of $M_{\rm
C}: M_{\rm O}: M_{\rm Ne}: M_{\rm Mg}: M_{\rm Si}: M_{\rm S}: M_{\rm
Ca}: M_{\rm Fe}= 0.006: 0.71: 0.037: 0.034: 0.083: 0.041: 0.007 :
0.09$. In the following calculation, we will use this composition as
a representative case for cosmic rays from the hypernova ejecta.
{ In both scenarios,  the amount of  hydrogen is negligible,
because the progenitor of Type Ib/c SN is stripped of H envelope, as
indicated by lack of H line in the SN spectrum (Iwamoto et al. 1998;
Nakamura et al. 2001). So during the explosion stage, there is
little amount of H in both the wind and the SN ejecta.}

\section{The propagated spectrum and chemical composition}
\subsection{Propagation of UHE CR nuclei}
While UHECRs are propagating in the intergalactic space, there are
three main attenuation processes  due to interaction with CMB or CIB
photons: the photodisintegration process, the Bethe--Heitler process
and the photopion production process. The photodisintegration
process causes nuclei  to lose their nucleons and thus change their
species, but does not reduce their Lorentz factor, while the latter
two processes can reduce the Lorentz factors of nuclei.  Thus,  we
need to consider the evolution of the mass number $A(t)$ and Lorentz
factor $\gamma_N(t)$ of a nucleus with time jointly. In the
hypernovae scenario, given that $E_{\rm max}/A \la 10^{19}$eV, the
corresponding maximum energy of each nucleon is below the threshold
energy for photopion production with CMB photons even at high
redshifts. Although these nuclei can interact with more energetic
CIB photons,  the photopion energy loss rate with such photons is
much lower than the energy loss via the Bethe--Heitler process by
interacting with much denser CMB photons, so we neglect the
photopion energy loss in the following calculation. In this work, we
adopt the CIB model suggested by \citet{Razzaque09} and
\citet{Finke10}, and use the online data  of the CIB intensity at
different redshifts
\footnote{http://www.phy.ohiou.edu/$\sim$finke/EBL/index.html}.

When an UHE nucleus is propagating in the intergalactic space, its
Lorentz factor evolves with time as
\begin{equation}\label{Et}
-\frac{d\gamma_N(t)}{dt}=\gamma_N(t)H(z)+\dot{\gamma}_{N,\rm BH}(t)
\end{equation}
where $\gamma_N(t)H(z)$ represents the adiabatic energy losses due
to cosmological expansion and $\dot{\gamma}_{N,\rm BH}(t)$ is energy
loss rate due to the Bethe-Heitler process. Here
$H(z)=H_0\sqrt{\Omega_m(1+z)^3+\Omega_{\Lambda}}$ is  the Hubble
constant at time $t=t(z)$, where $H_0=71\rm kms^{-1}Mpc^{-1}$,
$\Omega_m=0.27$ and $\Omega_{\Lambda}=0.73$. In an isotropic photon
background, the Bethe-Heitler energy loss rate of a nucleus of the
Lorentz factor $\gamma_i$ is given by
\begin{equation}
\begin{split}
\dot{\gamma}_{N,\rm BH}(t)=&\frac{Z^2}{A}\dot{\gamma}_{p,\rm BH}(t)\\
=&\frac{Z^2}{A}\frac{c}{2\gamma_N}\int_{\epsilon_{\rm
th}}^{\infty}d\epsilon_{\gamma}\sigma_{\rm
BH}(\epsilon_{\gamma})f_{\rm
BH}(\epsilon_{\gamma})\epsilon_{\gamma}\int_{\epsilon_{\gamma}/2\gamma_i}^{\infty}d\epsilon\frac{n_{\gamma}(\epsilon
,z)}{\epsilon^2}
\end{split}
\end{equation}
where $\dot{\gamma}_{p,\rm BH}(t)$ is the Bethe--Heitler energy loss
rate for a proton  of the same Lorentz factor $\gamma_N$,  $Z$ is
the nuclear charge of the nucleus, $\sigma_{\rm BH}$ and $f_{\rm
BH}$ are the cross section and the fraction of energy loss in one
interaction for the Beth-Heitler process \citep{Chodorowski92}, and
$\epsilon_{\rm th}$ is the threshold energy for the interaction.
Here $\epsilon_{\gamma}$ is the photon energy in the rest frame of
the nucleus while $\epsilon$ is the photon energy in the lab frame.
$n_{\gamma}(\epsilon,z)$ is the number density of background
(CMB+CIB) photons of energy $\epsilon$ at redshift $z$.

The nucleus suffers from loss of nucleons by the
photo-disintegration process, so the mass number  evolves as
\begin{equation}\label{At}
-\frac{dA}{dt}=R_A(t,\gamma_i),
\end{equation}
where $R_A$ is the reaction rate, which is
\begin{equation}
R_A(t,\gamma_N)=\frac{c}{2\gamma_N^2}\int_{\epsilon_{\rm
th}}^{\infty}d\epsilon \sigma_{{\rm dis},A}
(\epsilon)\epsilon\int_{\epsilon/2\gamma_i}^{\infty}d\epsilon_{\gamma}\frac{n_{\gamma}(\epsilon_{\gamma},z)}{\epsilon_{\gamma}^2},
\end{equation}
where $\sigma_{{\rm dis},A}$ is the total photo-disintegration cross
section for a nucleus of mass number $A$.  The photo-disintegration
cross section is dominated by the giant dipole resonance (GDR) up to
30MeV, with the threshold energy between about 10MeV and 20MeV for
all nuclei (in the nucleus rest frame). The cross section in this
energy range can be modeled by a Gaussian form (e.g.
\citealt{Puget76,Stecker99,Hooper08}) or a Lorentzian form (e.g.
\citealt{Khan05,Anchordoqui07,Hooper07}). From 30MeV to 150MeV, the
quasi-deuteron process becomes dominant and the cross section can be
approximated as  a plateau (e.g.
\citealt{Puget76,Stecker99,Ahlers10}). In this work, we use the
tabulated cross section data  generated by the code TALYS
\footnote{http://www.talys.edu/}\citep{Goriely08} which considered
all the individual nucleon emission channels for nuclei with $A\geq
5$. For nuclei with $A < 5$, we adopt the Gaussian form near the
threshold along with a plateau at higher energies to describe the
total cross section.

{We use the  Runge--Kutta method to solve Eq.~\ref{Et} and
Eq.~\ref{At} jointly so that  we can trace the evolution history of
the Lorentz factor $\gamma_N(z;z_s,E_s,A_s)$ and mass number
$A(z;z_s,E_s,A_s)$ of a specific nucleus injected from the source at
redshift $z=z_s$, where $E_s$ and $A_s$ are the initial energy and
mass number of the nucleus. The evolution of the mass number is
assumed to develop along the Puget-Stecker-Bredekamp chain (see e.g.
\citealt{Puget76, Stecker99, Ahlers10}). We also record the
redshifts,  Lorentz factors and number of secondary
protons\footnote{Since the neutrons emitted by the
photodisintegration process will decay into protons very soon
compared to the propagation time, we just treat the neutrons emitted
as protons of the same energy in the calculation.} that are
disintegrated from the parent nuclei so that the contribution of
secondary protons to cosmic-ray spectrum can  be properly included.}
The nuclei are injected from the maximum redshift $z_{\rm max}=6$ to
the minimum redshift $z_{\rm min}=0.001$ (corresponding to a
distance $D=4$ Mpc). The number density of hypernovae at redshift
$z$ is assumed to follow the star formation history (SFH), given by
\citep{Hopkins06, Yuksel08}
\begin{equation}
n(z)\propto \left \{
\begin{array}{lll}
(1+z)^{3.4}, z<1 \\
(1+z)^{-0.3}, 1<z<4 \\
(1+z)^{-3.5}, z>4 .
\end{array}
\right .
\end{equation}
We collect all the nuclei and secondary products that arrive at the
Earth ($z=0$) and put them into the corresponding energy bin and
species bin according to their Lorentz factors $\gamma(z=0)$ and
mass numbers $A(z=0)$. Then we can get the energy spectrum of cosmic
rays in each species bin, or  get the all-particle energy spectrum
by adding up all species particles at a certain energy.

\subsection{The final spectrum and composition}
The cosmic ray composition at the sources strongly affect the final
composition and energy spectrum of cosmic rays arriving at the
Earth. First, we consider the chemical composition at the source
equal to that of the WR stellar wind, where the mass ratio among
dominant elements is $M_{\rm He}: M_{\rm C}: M_{\rm O}=0.32: 0.39:
0.25$ with negligible H. The initial spectrum at the source is
assumed to be a power law with $p=2$. The result of the final
spectrum and  composition are presented in Fig.~1. One can see that
with the typical WR stellar wind composition, the cosmic ray flux
drops too fast at high energies  to account for the observations by
PAO. The apparent reason for the fast drop is that the low $E_{\rm
max}$ for intermediate mass nuclei such as C and O results in an
exponentially cutoff at energy $\la10^{20}$eV. However, the
situation will not be ameliorated even if we  raise $E_{\rm max,O}$
to higher energies, e.g. to $3\times 10^{20}$eV. { This is
because that lower energy background photons, which are more
abundant, are involved in the photodisintegration interaction for
intermediate mass nuclei due to their higher Lorentz factors,
compared with heavy nuclei of the same energy}. Therefore the attenuation
lengths for the intermediate nuclei with energies larger than
$10^{20}$ eV are so small \citep[e.g.][]{Puget76,Allard06} that even
those who emitted by the nearest sources at $z_{\rm min}$ = 0.001
(i.e. 4Mpc) will be effectively attenuated by the background photons
and hence the flux of cosmic rays at such energies is severely
suppressed after propagation. Only if $E_{\rm max, O}$ is extremely
high (e.g. $\sim3\times10^{21}$eV) so that secondary protons of
energies up to $2-3\times 10^{20}$eV can be produced in
photodisintegration process, and then the spectrum becomes flat at
the highest energy end \citep{Allard07}. However, such a high
$E_{\rm max, O}$ can hardly be achieved in the hypernova scenario.

Presence of heavier nuclei in the source composition would be
beneficial to fit the observation data, since heavy nuclei have
higher steepening energies  due to interactions with background
photons. Hypernova ejecta produced after the explosion can provide
heavy elements besides intermediate mass elements. Fig.~2 shows the
propagated spectrum for cosmic ray composition at the sources equal
to the hypernova ejecta composition  in the "CO138E50" model for SN
1998bw \citep{Nakamura01}. With the contribution by heavy nuclei at
the sources, the propagated spectrum is in better agreement with the
PAO data than the wind composition case.

The fit can be further improved if  some non-standard effects are
taken into account. In panel a of Fig.~\ref{fig3}, we increase the
initial iron abundance by a factor of 3, i.e. iron nuclei constitute
a fraction of $\sim 25\%$ of the total mass at the sources. This
could be achieved since the enrichment of heavy elements is not
uniformly distributed spatially, as already seen in some Galactic
supernova remnants \citep[e.g.][]{Hwang00}. A larger explosion
energy would also lead to more heavy elements synthesized during the
explosion \citep[e.g.][]{Metzger11}. As iron nuclei have larger steepening energy in the
spectrum due to interactions with background photons, a harder
spectrum is expected at the highest energy if more iron nuclei are
present. This is consistent with earlier results in
\citet{Allard08}. Another possible effect is a hard injection CR
spectrum. Although in conventional shock acceleration theory, the
power-law index is $p\simeq2$ , some observations
\citep[e.g.][]{Horns04, Green09, Reynolds11, Abdo11} and theoretical
calculations
\citep[e.g.][]{Ellison96,Malkov97,Vainio99,Vainio03,Tammi09} have
suggested harder cosmic-ray spectra with indices $p<2$. In panel
$b$, we show the propagated spectrum for an  injection spectrum with
power-law index $p=1.6$.  One can see that the fit to the data gets
better than the case of $p=2$. Besides, the source density
distribution in the nearby universe could be nonuniform. Since our
Galaxy locates inside the Local group and Local supercluster, we
examine the effect of local overdensity  of  sources on the
propagated spectrum. Panel $c$ shows the  spectrum for a local
overdensity in the source number by a factor of 2 relative to the
average within the size of 30 Mpc \citep{Blanton01}. Due to the attenuation by
background photons, higher energy cosmic rays are mainly contributed
by closer sources. Thus a local excess of the source number density
increases the flux of higher energy cosmic rays and thus hardens the
propagated spectrum at the highest energy end.

We should note that, although the deviation between the propagated
spectrum and observational data is relatively large in the wind
composition scenario, the situation can also be improved if some of
the above effects are taken into account. Panel d of Fig.~\ref{fig3}
presents the result after considering a harder injection cosmic-ray
spectrum of $p=-1.6$ as well as a local overdensity in the source
number by a factor of 2. The theoretical energy spectrum agrees
reasonably well with the observational data.

In Fig.~\ref{fig4}, we present how the average mass number of cosmic
rays evolves with energy for different scenarios discussed above.
{Note that the starting energy of this figure is set to $3\times
10^{18}$eV since our model do not account for the CR spectrum data
below this energy, for which a Galactic component contribution is
needed}. One can see that, for the hypernova ejecta composition
scenarios, the average mass number increases gradually with energy,
which is consistent with the finding by PAO that the composition of
UHECRs becomes increasingly heavy with energy. In the wind
composition scenarios, the average mass number increases more slowly
with energy. However, since the hadronic interaction models at such
high energies are not well understood, such a composition is still
consistent with the measurements within the uncertainties of
theoretical expectations.

\section{Discussions and Conclusions}
We have shown that, with some fraction of heavy nuclei such as iron
in the source composition, the semi-relativistic hypernova model can
explain the energy spectrum and composition of ultra-high energy
cosmic rays as measured by PAO. The heavy elements may originate
from the hypernova ejecta and  are brought into the forward shock
region via the Rayleigh-Taylor (R-T) instability. R-T fingers are
seen to be able to reach the forward shock front or overtake the
forward shocks \citep[e.g.][]{Jun96, BE01, Blondin01,Rakowski11}, so
they should be able to bring heavy elements into the ambient medium.
Observations of some Galactic supernova remnants such as Cas A and SN 1006,
{ as well as SN 1987A in Large Magellanic Cloud}, have provided direct and indirect evidences for
protrusion of heavy elements in the material  beyond the blast wave
\citep[e.g.][]{Erickson88,Fesen96,Hwang98,Hughes00}.

When the ejecta is being decelerated by the swept-up ambient wind
material, reverse shocks are also forming. The reverse shock may
accelerate particles in the shocked ejecta, which provides an
alternative site where UHE heavy nuclei are accelerated. The
reverser shock in the hypernova case is typically mildly
relativistic with  a speed about $\beta_{\rm rs}\simeq 0.5$,  as
shown in the appendix. Since the internal energy density in the
shocked ejecta and shocked wind are equal, the magnetic field in
the reverse shock is $B_{\rm rs}=(\epsilon_{B\rm
r}/\epsilon_B)^{1/2}B$, where $\epsilon_{B\rm r}$ is the magnetic
field energy equipartition factor in reverse shock and $\epsilon_B$
is the same factor for the forward shock. Then we obtain the maximum
energy of particles accelerated by the reverse shock
\begin{equation}
\begin{split}
E_{\rm max, rs}&=ZeB_{\rm rs}R_{\rm HN}\beta_{\rm rs}\\
&=1.8\times 10^{20} \left(\frac{Z}{26}\right)\epsilon_{B\rm
r,-1}^{1/2} \left(\frac{\beta_{\rm sh}\beta_{\rm
rs}}{0.5}\right)\dot{M}_{-5}^{1/2} v_{w,3}^{-1/2}{\rm eV},
\end{split}
\end{equation}
which can  also reach ultra-high energies for heavy  nuclei.

In summary, hypernovae remnants are shown to be potential sources of
UHECRs. They have been discovered to be present within the GZK
horizon, $\sim 100$ Mpc. Due to the semi-relativistic ejecta
produced by the explosion, they can accelerate particles to
ultra-high energies. Hypernovae  naturally provide intermediate mass
or heavy elements either in the form of stellar wind or the
hypernova ejecta itself. In this paper, we have calculated the
propagated spectrum and composition of UHECRs arriving at the Earth
given an initial cosmic ray spectrum and composition at the sources
produced by hyerpnovae. We find that with the source cosmic-ray
composition equal to that of the hypernova ejecta, the propagated
energy spectrum and composition are compatible with the measurements
by PAO.

{ There are suggestions that past Galactic hypernova could
contribute to cosmic rays in the energy range above the knee
\citep{Budnik08} or even UHECRs \citep{Calvez10}. However,
the event rate of such Galactic hypernovae is largely unknown. There
has been evidence indicating that  hypernovae associated with
sub-energetic GRBs are preferentially found in low-metallicity
galaxies \citep[e.g.][]{Woosley06, Stanek06}, which would imply that
hypernova rates in normal metallicity galaxies such as our Milky Way
may be low.}

\acknowledgments  We are grateful to the anonymous referee for the
valuable report, and to A. M. Taylor, P. M\'esz\'aros, S. Razzaque
and Z. G. Dai for useful discussions. This work is supported by the
NSFC under grants 10973008 and 11033002, the 973 program under grant
2009CB824800, the program of NCET, and the Fok Ying Tung Education
Foundation

\vskip 1cm
\appendix

\section{The speed of reverse shocks}

When the ejecta is being decelerated by the swept-up ambient wind
material, reverse shocks are also forming. The shocked wind and
shocked ejecta are in pressure balance and separated by the contact
discontinuity. The semi-relativistic ejecta get decelerated at
$R_{\rm HN}\simeq 5\times10^{16} E_{k,51}
(\Gamma_0/2)^{-2}\dot{M}_{-5}^{-1} v_{w,3} {\rm cm}$, where $E_{k}$
and $\Gamma_0$ are, respectively, the total kinetic energy and the
initial Lorentz factor of the semi-relativistic ejecta. Since the
shock expansion is expected to be adiabatic,  we have
\begin{equation}\label{EC}
E_{\rm sh,w}+E_{\rm sh,ej}=E_{k},
\end{equation}
where $E_{\rm sh,w}\simeq
4/3\pi\sigma\beta^2\Gamma^2R^3\rho_w(R)c^2$ \citep{BM76} is the
total energy {transferred to the shocked wind from the blast wave},
$E_{\rm sh,ej}\simeq \Gamma^2(e'+\rho'c^2+\beta^2p')V_{\rm sh,ej}$
is the total energy in the shocked ejecta, $\Gamma$ is the Lorentz
factor of the shocked material and $\rho_w(R)$ is the wind density
at radius $R$. Here $\sigma$ is a coefficient with
$\sigma\rightarrow 0.35$ when $\beta\rightarrow 1$ and
$\sigma\rightarrow 0.73$ when $\beta\rightarrow 0$. We use
$\sigma=0.73-0.38\beta$ to bridge the gap between the
non-relativistic case and ultra-relativistic case, as used in
\citet{HDL98}. $e'=\frac{\hat{\gamma}\gamma_{\rm
rs}+1}{\hat{\gamma}-1}(\gamma_{\rm rs}-1)\rho_{\rm ej}c^2$,
$\rho'=\frac{\hat{\gamma}\gamma_{\rm rs}+1}{\hat{\gamma}-1}\rho_{\rm
ej}$ and $p'=(\hat{\gamma}-1)e'$ are respectively, the internal
energy density, rest mass density and pressure in the comoving frame
of shocked material, and $\hat{\gamma}$ is the adiabatic index.
$V_{\rm sh,ej}=V_0/(\frac{\hat{\gamma}\gamma_{\rm
rs}+1}{\hat{\gamma}-1})$ is the volume of shocked ejecta, where
$V_0$ is the volume of the ejecta before being shocked. Here
$\gamma_{\rm rs}$ is the Lorentz factor of the reverse shock, which
relates with the Lorentz factor $\Gamma$ of the shocked material by
$\gamma_{\rm rs}\simeq \Gamma_0\Gamma(1-\beta_0\beta_{\rm se})$,
where $\beta_0$ and $\beta_{\rm se}$ are the velocities of the
unshocked ejecta  and shocked ejecta respectively. Solving Eq.~(\ref{EC})
with typical values for the parameters, we finally get the Lorentz
factor and speed of the reverse shock,
\begin{equation}
\gamma\approx 1.15,  \beta_{\rm rs}=\sqrt{1-1/\gamma^2}\approx 0.5 .
\end{equation}

\clearpage

\begin{figure}
\plotone{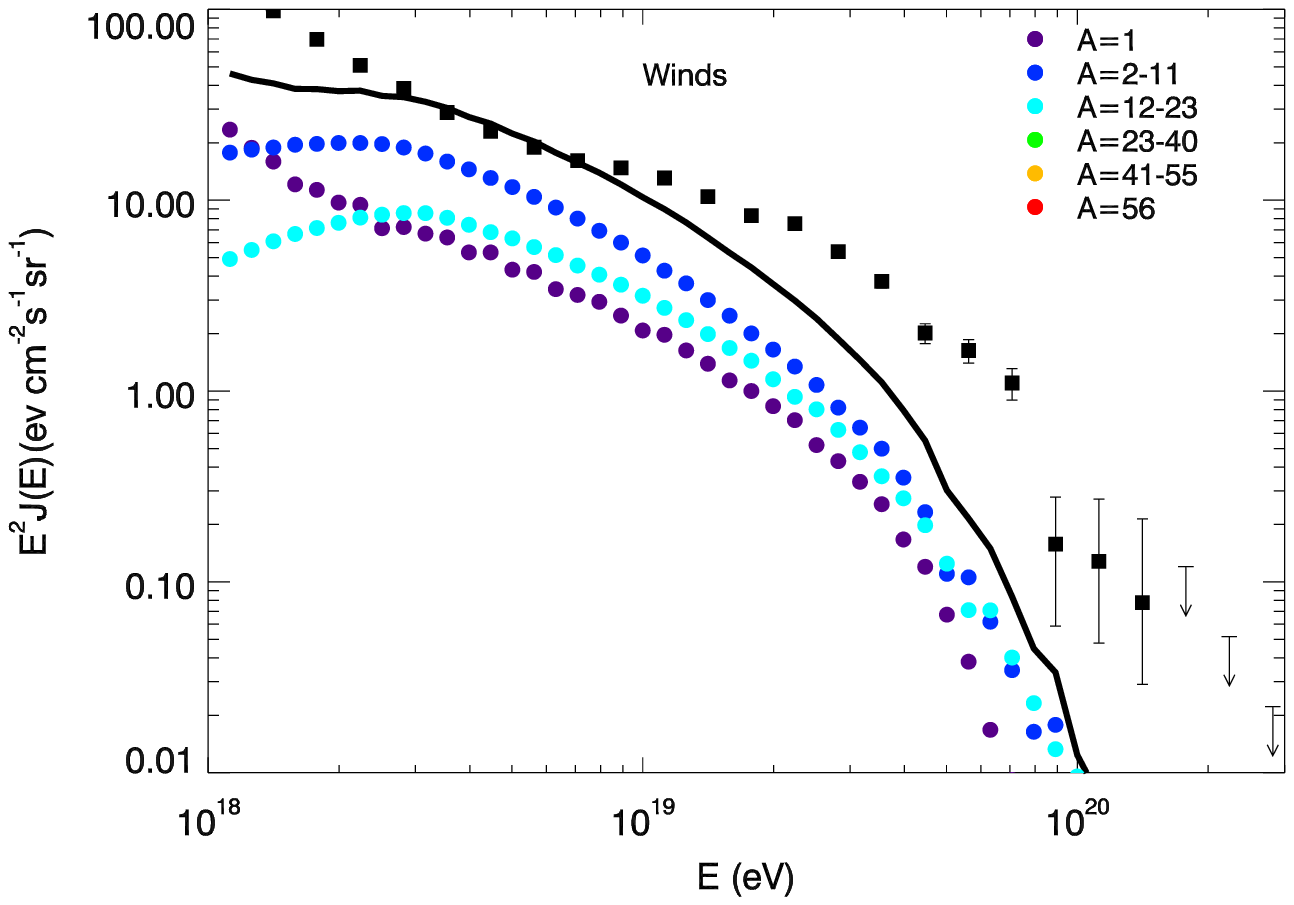} \caption{The propagated spectrum of cosmic rays
assuming a source composition equal to that of the stellar wind of
Wolf-Rayet stars, where the mass ratios of dominant elements are He:
C: O=0.32: 0.39: 0.25, and a flat source spectrum with $p=2$ (see
text for more details). The black solid line represents the
all-particle flux, while other lines represent the contributions by
different species as shown in the legend. The black squares are the
most recent data measured by PAO, taken from
http://www.auger.org/technical\_info/ICRC2011/CombinedSpectrum\_Auger2011.txt.\label{fig1}}
\end{figure}

\clearpage

\begin{figure}
\plotone{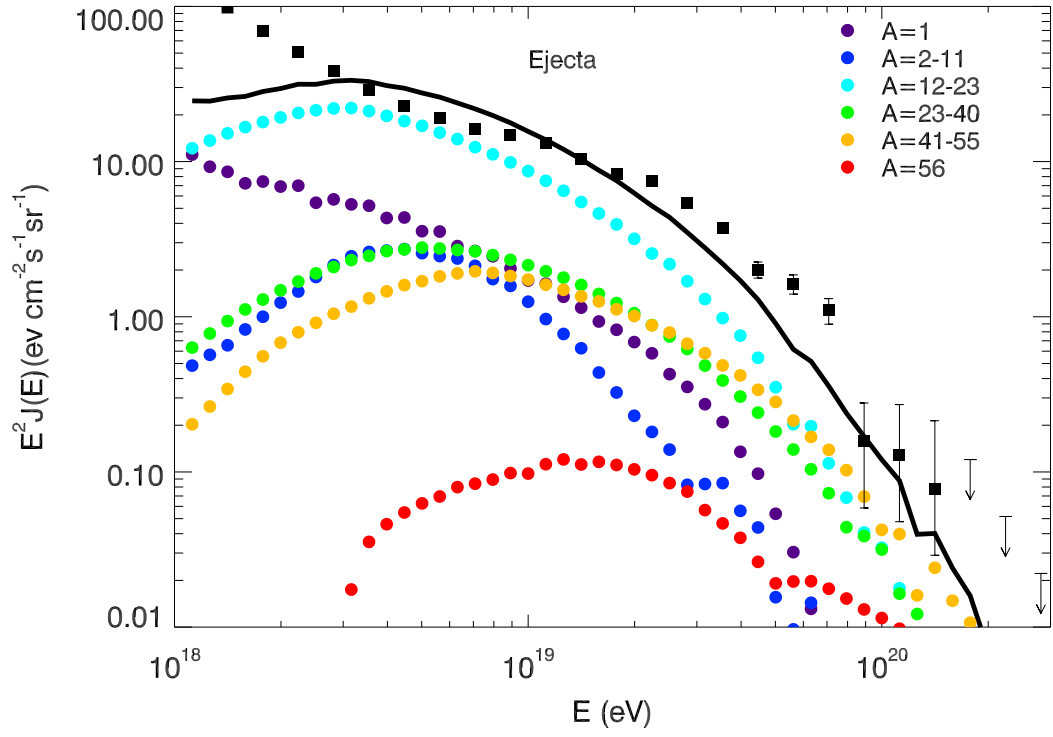} \caption{Same as Fig.1, but for a source
composition equal to that of the hypernova ejecta in the model of SN
1998bw (Nakamura et al. 2001). \label{fig2}}
\end{figure}

\clearpage

\begin{figure}
\plotone{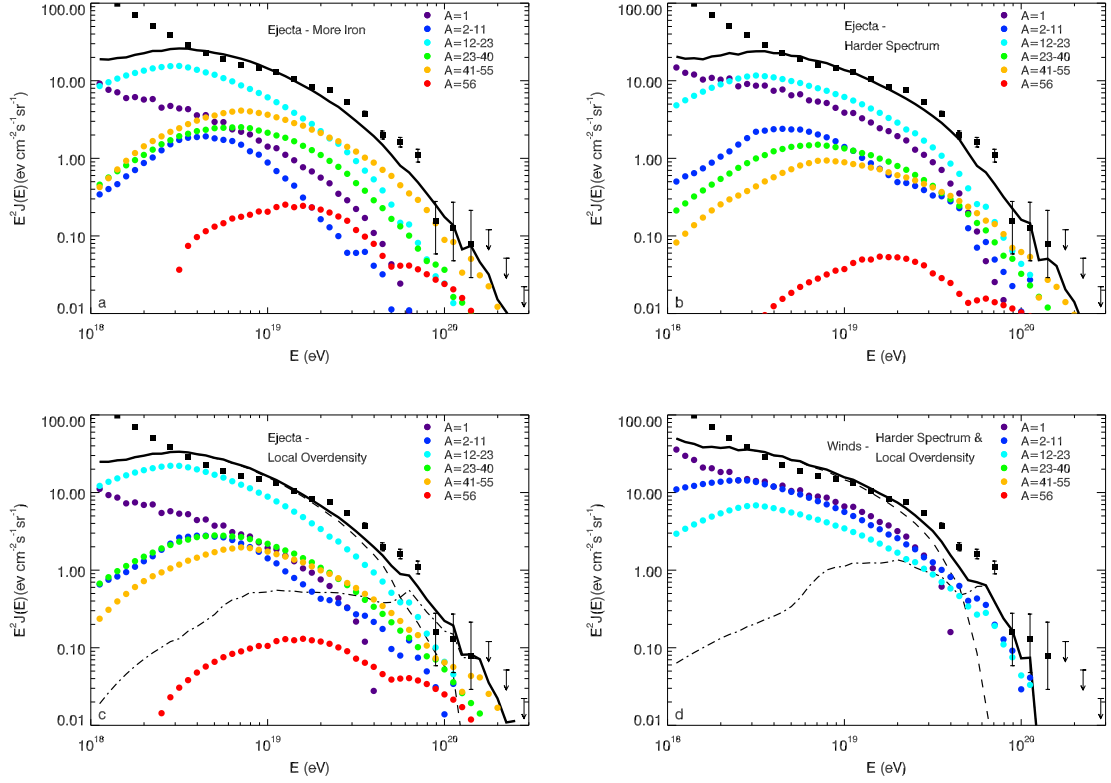} \caption{Panels $a,b,c$ show the propagated
cosmic ray spectrum for the source composition equal to that of the
hypernova ejecta, similar to Fig.~2, but with some modifications
taken into account. In panel $a$, we increase the abundance of iron
nuclei in the ejecta by a factor of 3. In panel $b$, we set a hard
initial spectrum with $p=1.6$. In panel $c$, we assume that the
local source number density is two times higher than the average
within a distance of  30 Mpc. The dashed line  in panel c represents
the contribution by sources beyond 30 Mpc while the dash--dotted
line represents the contribution by local sources within 30 Mpc.
Panel $d$ shows the propagated cosmic ray spectrum for the source
composition equal to that of stellar wind, but with the effects of a
hard spectrum and a local overdensity being taken into account. See
text for more details.\label{fig3}}
\end{figure}

\clearpage

\begin{figure}
\plotone{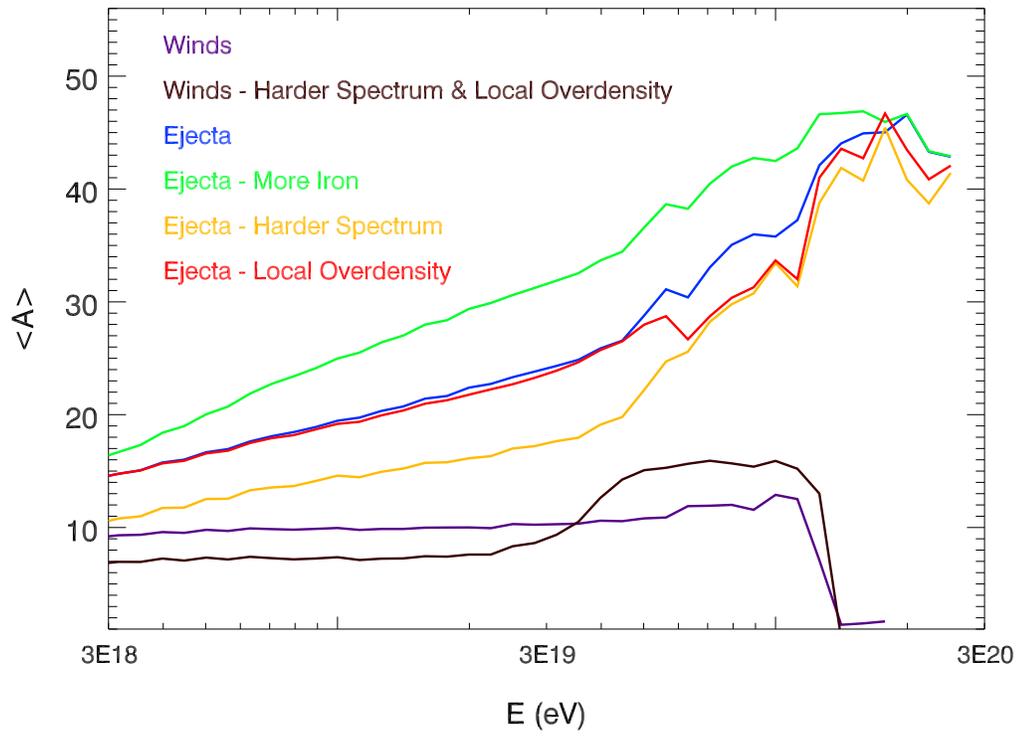} \caption{The average mass number of propagated
cosmic rays arriving at the Earth as a function of the cosmic-ray
energy for different cases shown in the above figures.
.\label{fig4}}
\end{figure}

\end{document}